\documentstyle[amssymb,aps,epsfig,twocolumn]{revtex}

\begin{document}
\title{Robust High-Fidelity Teleportation of an Atomic State through the Detection of
Cavity Decay}
\author{Bo Yu\thanks{%
leoncyu@mail.ustc.edu.cn}, Zheng-Wei Zhou, Yong Zhang, Guo-Yong Xiang and
Guang-Can Guo\thanks{%
gcguo@ustc.edu.cn}}
\address{Key Laboratory of Quantum Information, University of Science and Technology\\
of China, Hefei 230026, People's Republic of China}
\maketitle

\begin{abstract}
We propose a scheme for quantum teleportation of an atomic state based on
the detection of cavity decay. The internal state of an atom trapped in a
cavity can be disembodiedly transferred to another atom trapped in a distant
cavity by measuring interference of polarized photons through single-photon
detectors. In comparison with the original proposal by S. Bose, P.L. Knight,
M.B. Plenio, and V. Vedral [Phys. Rev. Lett. {\bf 83}, 5158 (1999)], our
protocol of teleportation has a high fidelity of almost unity, and inherent
robustness, such as the insensitivity of fidelity to randomness in the
atom's position, and to detection inefficiency. All these favorable features
make the scheme feasible with the current experimental technology.

{\bf PACS numbers:} 03.67.-a, 42.50.Gy, 32.80.Qk
\end{abstract}

\section{introduction}

Since the pioneering contribution of Bennett {\it et al.}\cite{1}{\it , }%
teleportation, which has recently attracted considerable attention as the
means of disembodied transfer of an unknown quantum state, comes to be
recognized as one of the basic methods of quantum communication and, more
generally, lies at the heart of the whole field of quantum information.
Experimental realizations of quantum teleportation have so far been focused
on the discrete-variables case, which involves photonic polarization states%
\cite{2,3} and vacuum--one-photon states\cite{4}, as well as the
continuous-variables one\cite{5}. However, since atoms are favorable for the
storage and processing of quantum information, teleportation of atomic
states will be the next important benchmark on the way to obtaining a
complete set of quantum information processing tools. Recently, a number of
proposals\cite{6,7} based on cavity quantum electrodynamics (QED) have been
presented for the teleportation of internal states of atoms. However, in the
earlier ones\cite{6}, atoms, which are not suited for long distance
transportation, have been used as flying qubits. From a practical point of
view, photons are the best candidate for flying qubits as the fast and
robust natural carriers of quantum information over long distance. In Ref.%
\cite{7}, Bose {\it et al. }designed a scheme for quantum teleportation with
a successful combination of the two advantages: atoms act as stationary
qubits, while photons play the role of flying qubits.

In this paper, we propose a scheme, which is similar to but more robust and
efficient than that of Bose {\it et al.}\cite{7}{\it ,} to teleport the
internal state of an atom trapped in a cavity onto a second atom trapped in
another distant cavity by detecting the photon decays from the cavities
through single-photon detectors. Instead of fighting against the decay of
the cavity field, which is seemed as a decoherence process resulting from
the unavoidable interaction of the cavity system with its surroundings, we
have designed an elegant scheme to use it as a constructive factor in the
teleportation of an atomic state. This kind of idea was widely discussed and
exploited very recently. Many schemes with this feature have been known for
entangling two or more atoms\cite{8,8'} as well as for entangling
macroscopic atomic ensembles\cite{8''}. Related protocols for quantum gate
operations and even universal quantum computation have also been proposed%
\cite{9}. Most of these schemes are based on the detection of cavity decay
and thus will succeed probabilistically only for particular measurement
results.

Although quantum information is similarly carried by photonic states, our
scheme is quite different from the previous quantum communication protocols%
\cite{10}. In the proposals in Ref.\cite{10}, quantum information is
directly transferred from an atom to another atom (both the atoms are
trapped in cavities) through a photon, thus the high requirement for the
experimental technology of feeding a photon into a cavity from the outside
must be fulfilled. However, in our scheme, the requirement is replaced by
detecting interference of polarized photons leaking out from both cavities,
which is highly developed and relatively simple to realize.

Compared with the original scheme\cite{7}, our protocol has some favorable
features such as robustness and high fidelity. Based on these helpful
advantages, which will be discussed in detail later, our scheme of
teleportation is expected to be implemented with the current cavity QED
experimental technology of trapping and manipulating single atoms\cite
{11,12,13}. Very recently, another similar teleportation scheme was proposed
by Cho and Lee\cite{14}. Both our scheme and that of Cho and Lee are based
on adiabatic processes, however, the pumping laser pulses in different
schemes are of different fashions. In Ref.\cite{14}, the atoms are driven by
$\pi $-polarized classical laser pulses which are perpendicular to the
cavity axes, whereas in our scheme, the driving laser pulses are kept
collinear with the cavity axes\cite{8'} and are circularly polarized. This
distinction makes the two schemes essentially different.

The paper is organized as follows: In Sec. II, in order to illustrate our
scheme explicitly, we first analyze the physical system of the scheme. In
Sec. III, we introduce the teleportation scheme in detail. A discussion on
the fidelity and advantages of our scheme is presented in Sec. IV. We
summarize the results in Sec. V.

\section{the physical system}

The framework of our proposal is schematically shown in Fig. 1. The system
we are considering consists of two optical cavities, with atoms 1 and 2
trapped in cavities {\it A} and {\it B,} respectively. The two atoms are
identical alkali atoms with different level structures involved, which are
composed of hyperfine and Zeeman sublevels\cite{15,16}. Both the atoms are
driven adiabatically through classical laser pulses which are collinear with
the cavity axes. Then the emitted photons, with quantum information carried
on the polarization states, will leak out from both of the cavities and
interfere at the device for Bell state measurement (BSM). Alice possesses
cavity {\it A}, atom 1 and BSM, and Bob holds cavity {\it B} and atom 2. The
whole procedure can be simply described as follow. Alice first maps her
atomic state to her polarization cavity state, while Bob, at the same time,
prepares a maximally entangled state of his atom and his polarization cavity
modes. Then all that Alice has to do is just to wait for the detection
result of the BSM device. Finally, Alice informs Bob of the detection result
via a classical communication channel, and Bob performs an appropriate local
unitary transformation to his atom to obtain the original teleported state.
With all these steps completed, Alice can efficiently teleport an unknown
internal state of her atom to that of Bob.

\begin{figure}[tbp]
\epsfig{file=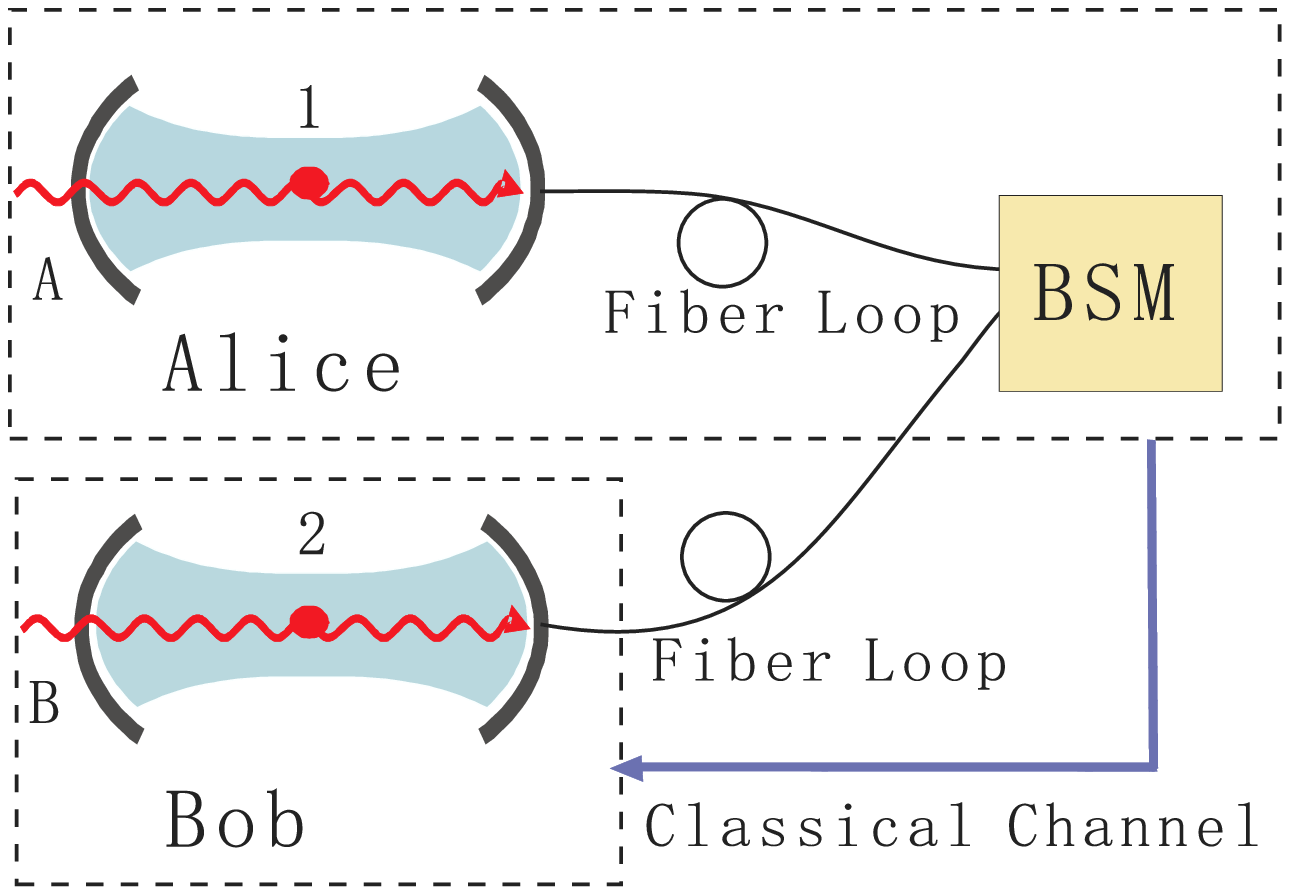,width=8cm}\caption{The schematic setup to
teleport the internal state of atom 1 trapped in cavity {\it A} to
that of atom 2 trapped in cavity {\it B}. BSM represents the
device for the Bell state measurement. The two fiber loops have
the same length.}
\end{figure}

In our proposal, the atoms are driven by classical laser pulses, which are
collinear with the cavity axes, through adiabatic passages. This new kind of
adiabatic scheme has been proposed by Duan {\it et al}. very recently\cite
{8'}. As we know, the coupling rate $g$ between the atomic internal levels
and the cavity mode depends on the atom position $\stackrel{\rightarrow }{r}$
through the relation $g(\stackrel{\rightarrow }{r})=C_gS(\stackrel{%
\rightarrow }{r})$, where $C_g$ is the corresponding Clebsch-Gordan (CG)
coefficient, and $S(\stackrel{\rightarrow }{r})$ is the spatial mode
function of the cavity mode with a definite constant $g_0$ incorporated.
Till now, most of the schemes based on high-Q optical cavities assumed that
the coupling rate $g$ is fixed. This assumption is tenable only when the
atom is localized to the Lamb-Dicke limit. However, it is still a bugbear in
experiment to satisfy the Lamb-Dicke condition, which prescribes that the
thermal oscillation amplitude of the atom must be small compared with the
optical wavelength. Therefore, an ingenious method is required to overcome
this experimental obstacle. If we keep the pumping laser incident from one
mirror of the cavity and collinear with the cavity axis, the classical
driving pulse has the same spatial mode structure as the cavity mode.
Accordingly, the Rabi frequency $\Omega $ between the atomic internal levels
and laser pulse can be similarly factorized as $\Omega (\stackrel{%
\rightarrow }{r},t)=C_\Omega S(\stackrel{\rightarrow }{r})\widetilde{E}(t)$,
where $C_\Omega $ is the corresponding CG coefficient (in the following, all
the $C_{g_i}$ and $C_{\Omega _j}$ are the corresponding CG coefficients),
and $\widetilde{E}(t)$ is proportional to the slowly-varying amplitude of
the driving pulse by a constant $R$. If an adiabatic evolution is
appropriately designed so that the relevant dynamics only depend on the
ratio $\Omega (\stackrel{\rightarrow }{r},t)/g(\stackrel{\rightarrow }{r})$,
which becomes independent of the random atom position $\stackrel{\rightarrow
}{r}$, it will go beyond the restriction of the Lamb-Dicke condition.

\begin{figure}[tbp]
\epsfig{file=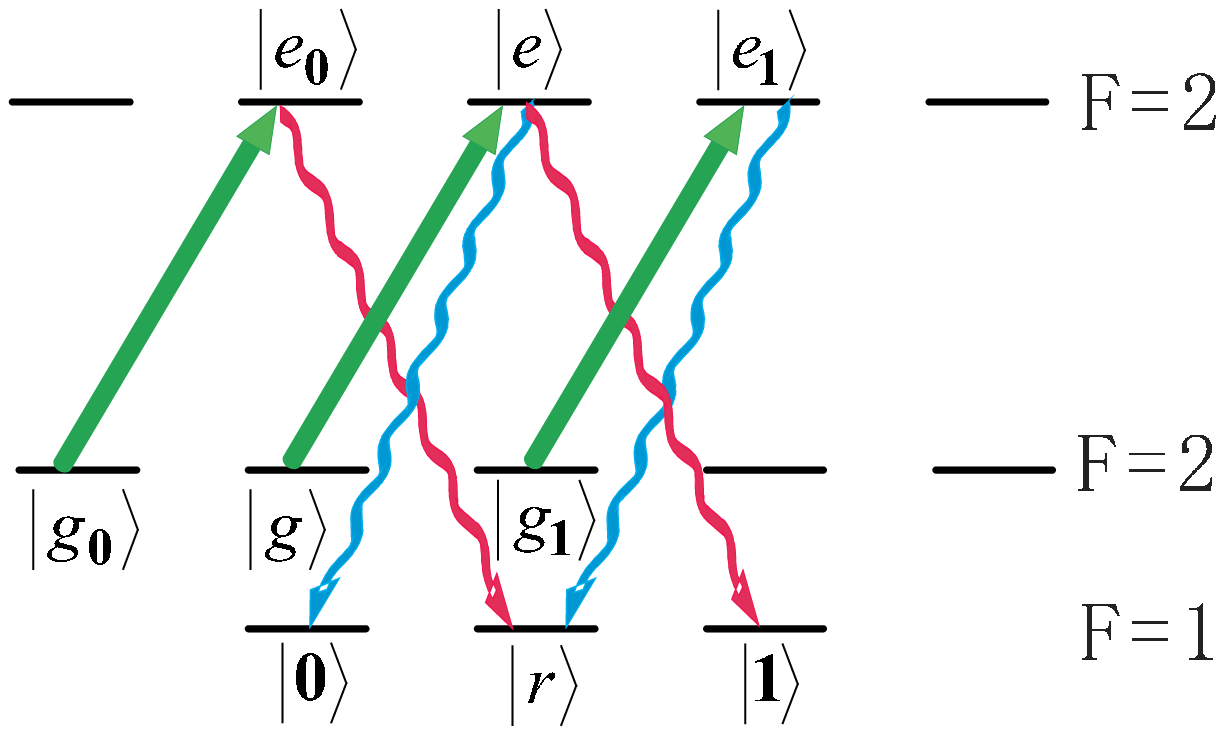,width=8cm}\caption{The relevant level
structures and transitions of atom 1 and atom 2. The two atoms are
identical alkali atoms, for example ${}^{87}$Rb, but involve
different atomic levels. Atom 1 exploits $\left| g_0\right\rangle
$, $\left| g_1\right\rangle $, $\left| e_0\right\rangle $, $\left|
e_1\right\rangle $, and $\left| r\right\rangle $, while atom 2 exploits $%
\left| g\right\rangle $, $\left| e\right\rangle $, $\left| 0\right\rangle $,
and $\left| 1\right\rangle $. The states $\left| g_0\right\rangle $, $\left|
g_1\right\rangle $, $\left| g\right\rangle $ ($\left| r\right\rangle $, $%
\left| 0\right\rangle $, $\left| 1\right\rangle $) correspond to the Zeeman
sublevels of the $F=2$ ($F=1$) ground hyperfine level, and $\left|
e_0\right\rangle $, $\left| e_1\right\rangle $, $\left| e\right\rangle $
correspond to the Zeeman sublevels of the $F=2$ excited hyperfine level.}
\end{figure}

The level structures of atom 1 and atom 2 are jointly shown in Fig. 2. Such
atomic level structures can be achieved in ${}^{87}$Rb, so we take $^{87}$Rb
as our choice. The states $\left| g_0\right\rangle $, $\left|
g_1\right\rangle $, $\left| g\right\rangle $, $\left| 0\right\rangle $, $%
\left| 1\right\rangle $, and $\left| r\right\rangle $ correspond to $\left|
F\right. =2,m=-\left. 2\right\rangle $, $\left| F\right. =2,m=\left.
0\right\rangle $, $\left| F\right. =2,m=-\left. 1\right\rangle $, $\left|
F\right. =1,m=-\left. 1\right\rangle $, $\left| F\right. =1,m=\left.
1\right\rangle $, and $\left| F\right. =1,m=\left. 0\right\rangle $ of $%
5S_{1/2}$, respectively. $\left| e_0\right\rangle $, $\left|
e_1\right\rangle $, and $\left| e\right\rangle $ correspond to $\left|
F\right. =2,m=-\left. 1\right\rangle $, $\left| F\right. =2,m=\left.
1\right\rangle $, and $\left| F\right. =2,m=\left. 0\right\rangle $ of $%
5P_{1/2}$, respectively. The qubit of Alice (atom 1) is encoded in $\left|
g_0\right\rangle $ and $\left| g_1\right\rangle $, while the qubit of Bob
(atom 2) is encoded in $\left| 0\right\rangle $ and $\left| 1\right\rangle $%
. The transitions $\left| g_0\right\rangle \rightarrow \left|
e_0\right\rangle $, $\left| g_1\right\rangle \rightarrow \left|
e_1\right\rangle $ and $\left| g\right\rangle \rightarrow \left|
e\right\rangle $ are driven resonantly and adiabatically by right-circularly
polarized classical laser pulses, with the corresponding Rabi frequencies
signified by $\Omega _0(t)$, $\Omega _1(t)$ and $\Omega _2(t)$ respectively.
The transitions $\left| e_0\right\rangle \rightarrow \left| r\right\rangle $
and $\left| e\right\rangle \rightarrow \left| 1\right\rangle $ ($\left|
e_1\right\rangle \rightarrow \left| r\right\rangle $ and $\left|
e\right\rangle \rightarrow \left| 0\right\rangle $) are resonantly coupled
to the cavity mode $a_L$ ($a_R$) with left-circularly (right-circularly)
polarization. Because of the symmetry of the atomic level structures, the
coupling rates corresponding to $\left| e_0\right\rangle \rightarrow \left|
r\right\rangle $ and $\left| e_1\right\rangle \rightarrow \left|
r\right\rangle $ can be simultaneously denoted by $g_1$, while those
corresponding to $\left| e\right\rangle \rightarrow \left| 0\right\rangle $
and $\left| e\right\rangle \rightarrow \left| 1\right\rangle $ can be
simultaneously denoted by $g_2$. Without loss of generality, all the Rabi
frequencies and coupling rates are assumed to be real.

\section{the teleportation scheme}

The arbitrary unknown state of atom 1 that is to be transferred from Alice
to Bob can be written as
\begin{equation}
\left| \psi \right\rangle _1=a\left| g_0\right\rangle _1+b\left|
g_1\right\rangle _1,  \eqnum{1}
\end{equation}
where $a$ and $b$ are complex probability amplitudes, and $\left| a\right|
^2+\left| b\right| ^2=1$. With cavity A prepared in the vacuum state $\left|
0\right\rangle _A$, the initial state of the whole system of Alice is $%
(a\left| g_0\right\rangle _1+b\left| g_1\right\rangle _1)\left|
0\right\rangle _A$. If the transitions $\left| g_0\right\rangle \rightarrow
\left| e_0\right\rangle $ and $\left| g_1\right\rangle \rightarrow \left|
e_1\right\rangle $ are driven adiabatically by laser pulse collinear with
the cavity axis, atom 1 will be transferred with probability $P_1\approx 1$
to the state $\left| r\right\rangle _1$ by emitting a photon from the
transition $\left| e_0\right\rangle \rightarrow \left| r\right\rangle $ or $%
\left| e_1\right\rangle \rightarrow \left| r\right\rangle $. The Hamiltonian
of Alice's system in the rotating frame is given by (assuming $\hbar =1$)
\begin{eqnarray}
H_1 &=&i\Omega _0(t)(A_0-A_0^{\dagger })+i\Omega _1(t)(A_1-A_1^{\dagger })-
\nonumber \\
&&ig_1(a_L^{A\dagger }A_L^A-A_L^{A\dagger }a_L^A)-ig_1(a_R^{A\dagger
}A_R^A-A_R^{A\dagger }a_R^A),  \eqnum{2}
\end{eqnarray}
where $A_0=\left| g_0\right\rangle _1\left\langle e_0\right| $, $A_1=\left|
g_1\right\rangle _1\left\langle e_1\right| $, $A_L^A=\left| r\right\rangle
_1\left\langle e_0\right| $, $A_R^A=\left| r\right\rangle _1\left\langle
e_1\right| $, and $a_L^A$ ($a_R^A$) represents the annihilation operator for
the left-circularly (right-circularly) polarized mode of cavity A. The
Hamiltonian $H_1$ has two orthogonal dark states:
\begin{equation}
\left| D_0\right\rangle =\frac{g_1\left| g_0\right\rangle _1\left|
0\right\rangle _A+\Omega _0(t)\left| r\right\rangle _1\left| L\right\rangle
_A}{\sqrt{g_1^2+\Omega _0^2(t)}},  \eqnum{3}
\end{equation}
and
\begin{equation}
\left| D_1\right\rangle =\frac{g_1\left| g_1\right\rangle _1\left|
0\right\rangle _A+\Omega _1(t)\left| r\right\rangle _1\left| R\right\rangle
_A}{\sqrt{g_1^2+\Omega _1^2(t)}}.  \eqnum{4}
\end{equation}
Under the adiabatic approximation, the state of Alice's system at time $t$
has the form
\begin{equation}
\left| \Psi (t)\right\rangle _{1A}=ae^{i\varphi _0(t)}\left|
D_0\right\rangle +be^{i\varphi _1(t)}\left| D_1\right\rangle ,  \eqnum{5}
\end{equation}
with
\begin{equation}
\varphi _k(t)=i\int_0^td\tau \left\langle D_k\right| \frac \partial {%
\partial \tau }\left| D_k\right\rangle -\int_0^td\tau E_k(\tau ),  \eqnum{6}
\end{equation}
where $k=0,1$, and $E_k(t)$ is the corresponding energy eigenvalue. Here, we
have $E_0(t)=E_1(t)=0$. In Equation (6), the first term on the right side is
the adiabatic phase or so-called Berry phase, and the second term is the
dynamical phase. Obviously, we have $\varphi _0(t)=\varphi _1(t)=0$, so $%
\left| \Psi (t)\right\rangle _{1A}$ becomes
\begin{eqnarray}
\left| \Psi (t)\right\rangle _{1A} &=&a\left| D_0\right\rangle +b\left|
D_1\right\rangle   \nonumber \\
&=&(a\cos \theta _0(t)\left| g_0\right\rangle _1+b\cos \theta _1(t)\left|
g_1\right\rangle _1)\left| 0\right\rangle _A  \nonumber \\
&&+\left| r\right\rangle _1(a\sin \theta _0(t)\left| L\right\rangle _A+b\sin
\theta _1(t)\left| R\right\rangle _A),  \eqnum{7}
\end{eqnarray}
with
\begin{equation}
\cos \theta _i(t)=g_1/\sqrt{g_1^2+\Omega _i^2(t)},  \eqnum{8}
\end{equation}
and
\begin{equation}
\sin \theta _i(t)=\Omega _i(t)/\sqrt{g_1^2+\Omega _i^2(t)},  \eqnum{9}
\end{equation}
where $i=0,1$. The initial state $(a\left| g_0\right\rangle _1+b\left|
g_1\right\rangle _1)\left| 0\right\rangle _A$ finally evolves into $\left|
r\right\rangle _1(a\left| L\right\rangle _A+b\left| R\right\rangle _A)$ with
$\Omega _i(t)$ increasing gradually.

At the same time, Bob switches on a similar laser pulse, which drives the
transition $\left| g\right\rangle \rightarrow \left| e\right\rangle $
adiabatically. With cavity B also prepared in the vacuum state $\left|
0\right\rangle _B$, atom 2, initially prepared in $\left| g\right\rangle _2$%
, will be transferred with probability $P_2\approx 1$ to the states $\left|
0\right\rangle _2$ and $\left| 1\right\rangle _2$ by emitting a photon from
the transition $\left| e\right\rangle \rightarrow \left| 0\right\rangle $ or
$\left| e\right\rangle \rightarrow \left| 1\right\rangle $. The Hamiltonian
of Bob's system in the rotating frame is given by
\begin{eqnarray}
H_2 &=&i\Omega _2(t)(A_2-A_2^{\dagger })-ig_2(a_L^{B\dagger
}A_L^B-A_L^{B\dagger }a_L^B)  \nonumber \\
&&-ig_2(a_R^{B\dagger }A_R^B-A_R^{B\dagger }a_R^B),  \eqnum{10}
\end{eqnarray}
where $A_2=\left| g\right\rangle _2\left\langle e\right| $, $A_L^B=\left|
1\right\rangle _2\left\langle e\right| $, $A_R^B=\left| 0\right\rangle
_2\left\langle e\right| $, and $a_L^B$ ($a_R^B$) represents the annihilation
operator for the left-circularly (right-circularly) polarized mode of cavity
B. The Hamiltonian $H_2$ has a dark state:
\begin{equation}
\left| D_2\right\rangle =\frac{\sqrt{2}g_2\left| g\right\rangle _2\left|
0\right\rangle _B+\Omega _2(t)\frac{\left| 0\right\rangle _2\left|
R\right\rangle _B+\left| 1\right\rangle _2\left| L\right\rangle _B}{\sqrt{2}}%
}{\sqrt{2g_2^2+\Omega _2^2(t)}}.  \eqnum{11}
\end{equation}
Under the adiabatic approximation, the state of Bob's system at time $t$ has
the form
\begin{eqnarray}
\left| \Phi (t)\right\rangle _{2B} &=&\cos \theta _2(t)\left| g\right\rangle
_2\left| 0\right\rangle _B  \nonumber \\
&&+\sin \theta _2(t)(\left| 0\right\rangle _2\left| R\right\rangle _B+\left|
1\right\rangle _2\left| L\right\rangle _B)/\sqrt{2},  \eqnum{12}
\end{eqnarray}
with
\begin{equation}
\cos \theta _2(t)=\sqrt{2}g_2/\sqrt{2g_2^2+\Omega _2^2(t)},  \eqnum{13}
\end{equation}
and
\begin{equation}
\sin \theta _2(t)=\Omega _2(t)/\sqrt{2g_2^2+\Omega _2^2(t)}.  \eqnum{14}
\end{equation}
The initial state $\left| 0\right\rangle _B\left| g\right\rangle _2$ finally
evolves into $(\left| 0\right\rangle _2\left| R\right\rangle _B+\left|
1\right\rangle _2\left| L\right\rangle _B)/\sqrt{2}$ with $\Omega _2(t)$
increasing gradually.

Because of the imperfection of the cavities, the two emitted photons will
leak out from them and interfere at the device for BSM. Although the
complete BSM has been realized successfully in experiment\cite{3}, it is
inefficient since nonlinear processes are involved. Several mostly used BSMs
are based on linear optical elements, and only succeed in $50\%$ or smaller
of all the cases. The BSM of our scheme, which has a success probability of
the upper bound $50\%$, is shown in Fig. 3 (see Ref.\cite{17}). A
straightforward analysis shows that, with the BSM successfully completed on $%
\left| r\right\rangle _1(a\left| L\right\rangle _A+b\left| R\right\rangle
_A)(\left| 0\right\rangle _2\left| R\right\rangle _B+\left| 1\right\rangle
_2\left| L\right\rangle _B)/\sqrt{2}$, which is the joint state of Alice's
and Bob's systems, the state of atom 2 becomes $a\left| 0\right\rangle _2\pm
b\left| 1\right\rangle _2$. Concretely, if $D_{1,4}$ or $D_{2,3}$ ($D_{1,3}$
or $D_{2,4}$) are triggered, atom 2 will be on the state $a\left|
0\right\rangle _2+b\left| 1\right\rangle _2$ ($a\left| 0\right\rangle
_2-b\left| 1\right\rangle _2$). Otherwise, the teleportation fails. After
Alice has sent the result of the response of the detectors to Bob, he
performs an appropriate unitary operation on atom 2, and the teleportation
is thus finished.

\begin{figure}[tbp]
\epsfig{file=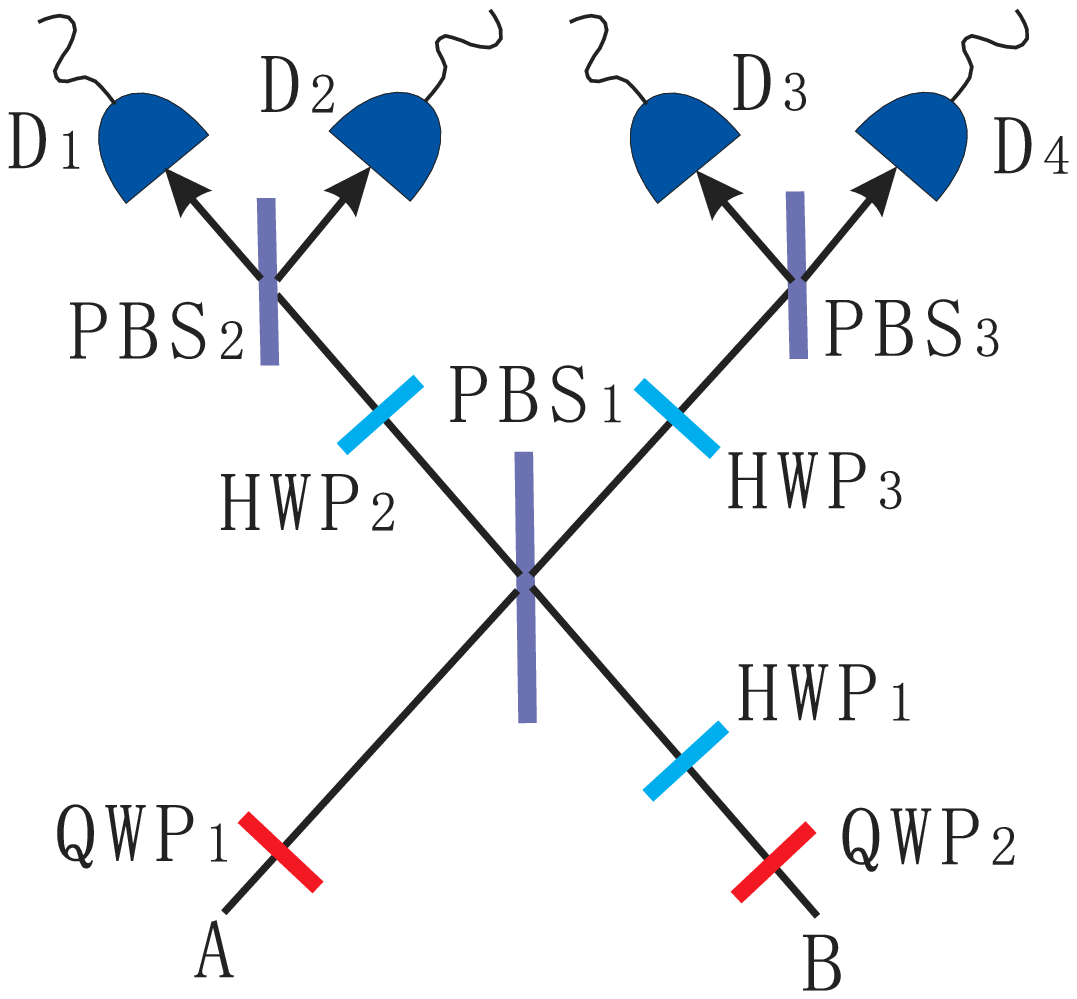,width=6cm}\caption{The device for Bell
state measurement. $PBS_{1,2,3}$ denote
polarization beam splitters, $QWP_{1,2}$ signify quarter wave plates, and $%
D_{1,2,3,4}$ represent detectors. $HWP_1$ is a $90^{\circ }$ half wave
plate, while $HWP_{2,3}$ are $45^{\circ }$ half wave plates.}
\end{figure}

In addition, we briefly consider the preparation of the initial state $%
a\left| g_0\right\rangle _1+b\left| g_1\right\rangle _1$ in the experimental
demonstration. In Ref.\cite{16}, a method is proposed by Law and Eberly to
prepare an arbitrarily prescribed superposition of internal Zeeman levels of
an atom by Raman pulses. If we apply this method, the initial state can be
easily generated. For example, we assume that atom 1 would be firstly
prepared in the state $\left| g_0\right\rangle _1$ by optical pumping. Fig.
4 shows the pulse sequence to generate the initial state $a\left|
g_0\right\rangle _1+b\left| g_1\right\rangle _1$. Step (1) forces the state $%
\left| g_0\right\rangle _1$ to evolve into $a\left| g_0\right\rangle
_1+b\left| 0\right\rangle _1$. In this step, the area of the Raman pulse
should be adjusted according to the CG coefficients and the complex
probability amplitudes $a$ and $b$. Step (2) completely transfers the
occupation of the state $\left| 0\right\rangle _1$ to that of the state $%
\left| g_1\right\rangle _1$, and thus the state of atom 1 becomes $a\left|
g_0\right\rangle _1+b\left| g_1\right\rangle _1$. In this step, the required
Raman pulse is a $\pi $ pulse.

\begin{figure}[tbp]
\epsfig{file=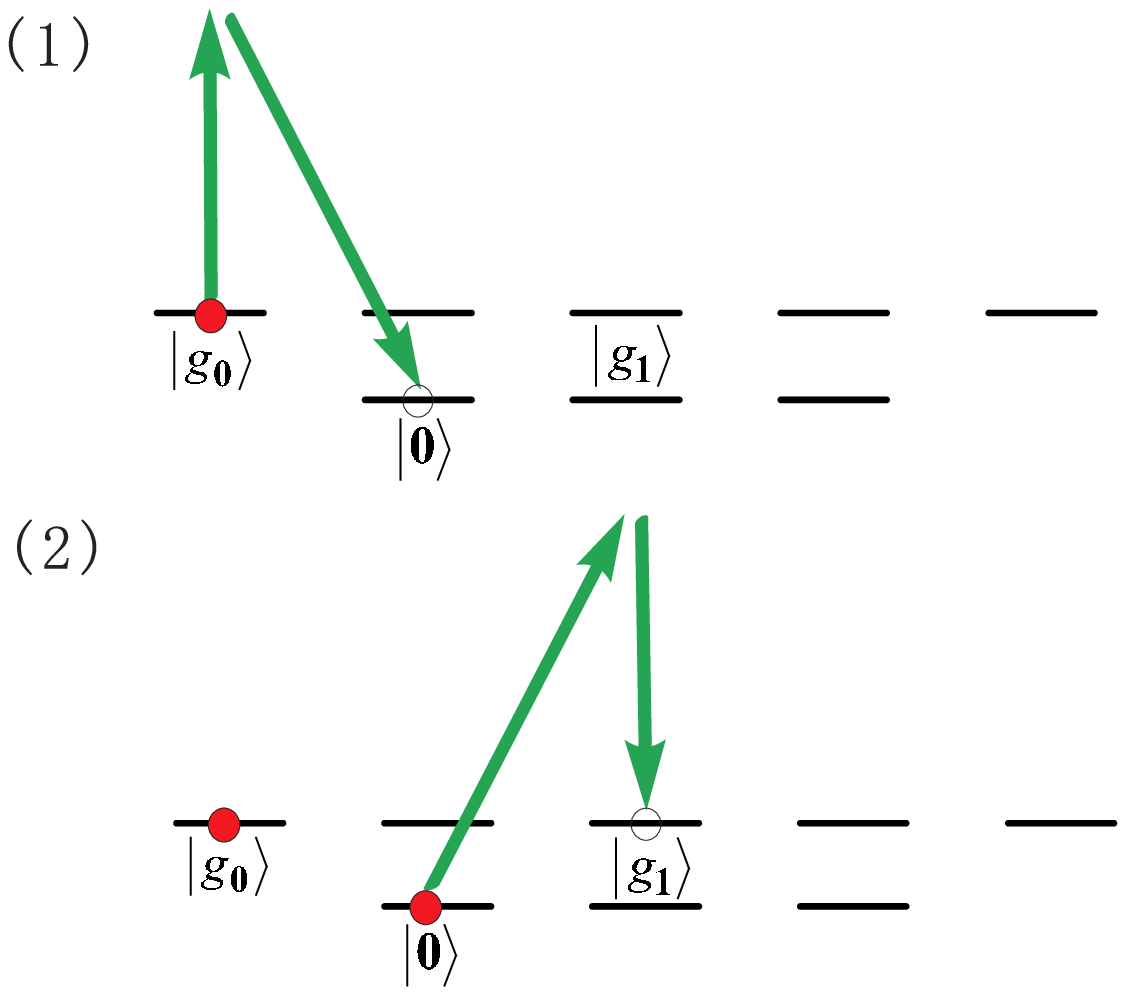,width=8cm}\caption{A 2-pulse sequence to
prepare the initial state $a\left| g_0\right\rangle _1+b\left|
g_1\right\rangle _1$. The solid circles represent the occupied
states, while the empty circles represent the states to be
occupied.}
\end{figure}

\section{discussion on the fidelity and advantages of the scheme}

Now we turn to the estimation of the fidelity. The BSM does a perfect job
only when the output pulse shapes of the two photons match exactly, however,
this condition can not be satisfied in our scheme. Approximately, the pulse
shape of Bob's photon\cite{8'} is analytically given by
\begin{equation}
f_B(t)=\sqrt{\kappa }\sin \theta _2(t)\exp [-(\kappa /2)\int_0^t\sin
^2\theta _2(\tau )d\tau ],  \eqnum{15}
\end{equation}
where $\kappa $ represents the common decay rate of cavity A and B. The
pulse shape of Alice's photon $f_A(t)$ alters with the initial state of atom
1. For special case $\left| \psi \right\rangle _1=\left| g_i\right\rangle _1$
($i=0,1$), we have
\begin{equation}
f_{Ai}(t)=\sqrt{\kappa }\sin \theta _i(t)\exp [-(\kappa /2)\int_0^t\sin
^2\theta _i(\tau )d\tau ],  \eqnum{16}
\end{equation}
where $f_{Ai}(t)$ is the corresponding pulse shape. For general case, $%
f_A(t) $ varies from $f_{A0}(t)$ to $f_{A1}(t)$. Therefore the fidelity of
our teleportation is highly determined by the difference $\delta $ between $%
f_{A0}(t)$ and $f_{A1}(t)$. Because
\begin{eqnarray}
\sin \theta _i(t) &=&\Omega _i(t)/\sqrt{g_1^2+\Omega _i^2(t)}  \nonumber \\
&=&C_{\Omega _i}\widetilde{E}_1(t)/\sqrt{C_{g_1}^2+C_{\Omega _i}^2\widetilde{%
E}_1^2(t)},  \eqnum{17}
\end{eqnarray}
where $\widetilde{E}_1(t)$ is proportional to the slowly-varying amplitude
of Alice's driving pulse by a constant $R$, $\delta $ is entirely generated
by the inequality of $C_{\Omega _i}$. Here, $C_{\Omega _0}=\sqrt{1/3}$ and $%
C_{\Omega _1}=\sqrt{1/2}$. Fortunately, if we choose an appropriate driving
pulse shape, $\delta $ can be small enough to be neglected. An example is
shown in Fig. 5, where, and in the following, the pulse shape functions are
renormalized according to $\int_0^Tf^2(t)dt=1$ ($T$ is the driving pulse
duration) for convenience of comparison. The two curves overlap very well,
and with $\delta $ quantified through $\delta
=1-\int_0^Tf_{A0}(t)f_{A1}(t)dt $, we obtain $1-\delta =0.992$. Because
\begin{eqnarray}
\sin \theta _2(t) &=&\Omega _2(t)/\sqrt{2g_2^2+\Omega _2^2(t)}  \nonumber \\
&=&C_{\Omega _2}\widetilde{E}_2(t)/\sqrt{2C_{g_2}^2+C_{\Omega _2}^2%
\widetilde{E}_2^2(t)},  \eqnum{18}
\end{eqnarray}
where $\widetilde{E}_2(t)$ is proportional to the slowly-varying amplitude
of Bob's driving pulse by a constant $R$, when $\widetilde{E}_2(t)$ is
chosen to satisfy
\begin{equation}
\widetilde{E}_2(t)=(\sqrt{2}C_{g_2}C_{\Omega _1}/C_{g_1}C_{\Omega _2})%
\widetilde{E}_1(t)=\sqrt{2/3}\widetilde{E}_1(t),  \eqnum{19}
\end{equation}
we have $f_B(t)=f_{A1}(t)$. The state-dependent fidelity $F$ of the final
state of atom 2 with respect to the initial state of atom 1 has the
following form
\begin{equation}
F=\sqrt{\left| a\right| ^4+\left| b\right| ^4+2\left| a\right| ^2\left|
b\right| ^2\left( \int_0^Tf_A(t)f_B(t)dt\right) ^2}.  \eqnum{20}
\end{equation}
Then it is straightforward that $F\gtrsim 1-\delta $ for arbitrary state, so
we almost have a fidelity of unity. Furthermore, as the inequality of $%
C_{\Omega _i}$ does not result in the large $\delta $ and thus the large
loss in the fidelity, our scheme has another favorable feature. Reasonable
as it seems, the fidelity is insensitive to the ratio $\widetilde{E}_2(t)/%
\widetilde{E}_1(t)$, with $\widetilde{E}_1(t)$ and $\widetilde{E}_2(t)$
sharing the same normalized driving pulse shape. So $\widetilde{E}_2(t)/%
\widetilde{E}_1(t)$ is not required to equal $\sqrt{2/3}$ accurately.

\begin{figure}[tbp]
\epsfig{file=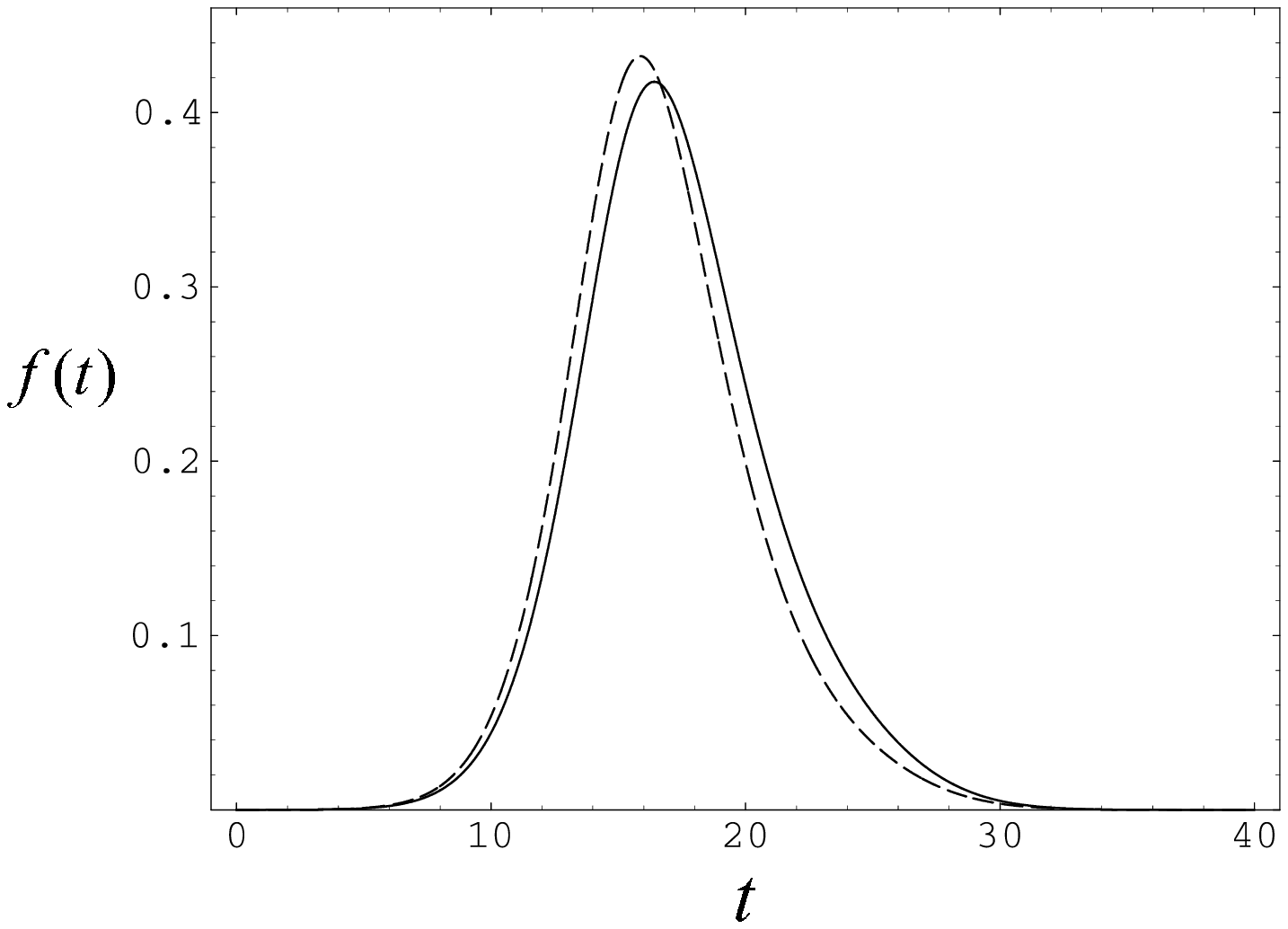,width=8cm}\caption{The pulse shape
functions $f_{A0}(t)$ (solid curve) and $f_{A1}(t)$ (dashed
curve). To satisfy the adiabatic condition, we have taken the
driving pulse duration $T=40/\kappa $. $\widetilde{E}_1(t)$ is in
a Gaussian shape with
the peak at $t=T/2$ and the width $t_w=\sqrt{2}T/10$, and $E_{1m}=C_{g_1}/$ $%
C_{\Omega _0}$, where $E_{1m}$ is the maximum of
$\widetilde{E}_1(t)$. }
\end{figure}

A presentation of the advantages of our scheme is now in order. First, our
scheme has a high fidelity, with a large success probability of $50\%$
achieved in the ideal case. As shown above, if the driving pulses are chosen
appropriately, the fidelity can be made higher than $0.99$ for arbitrary
state, and thus approaches unity approximately. Second, our scheme is
intrinsically robust to spontaneous emission. This atomic decay is highly
suppressed by the adiabatic method, and it only results in the loss of
photons even if it happens. Third, compared with the original scheme\cite{7}%
, our scheme also has inherent robustness to output coupling inefficiency of
the cavities, transmission loss, and detector inefficiency. In our scheme,
all these practical noises and technical imperfections only lead to the loss
of photons, and thus loss of the success probability, but have no influence
on the fidelity. Whereas in the original scheme, distinguishing between one
and two photons is required, so the decrease of the fidelity is inevitable.
Besides, the fidelity is insensitive to the ratio of the slowly-varying
amplitudes of the driving pulses $\widetilde{E}_2(t)/\widetilde{E}_1(t)$.
This feature removes the requirement to accurately control the intensity of
the laser pulse. Finally, our scheme successfully overcomes the experimental
difficulties caused by the randomness of the coupling rate. The Lamb-Dicke
condition is no longer needed to be satisfied. A far-off resonance trapping
(FORT) beam\cite{11,12,13} forms many potential wells along the cavity axis,
and the bottoms of different potential wells have different coupling rates.
In current experiments, one can not control and even does not know precisely
in which well the atom is trapped. But in our scheme this kind of
uncertainty of the coupling rate is well conquered. So our scheme of
teleportation is expected to be implemented with the current cavity QED
experimental technology of trapping and manipulating single atoms.

In comparison with the recent similar teleportation
scheme\cite{14}, our scheme is more robust to the randomness in
the atom's position. First, we do not need each atom to be trapped
in the same FORT potential well, and even we need no information
on which well the atom is trapped in. Second, the thermal
oscillation of the atom, which should be considered when the
Lamb-Dicke condition is not satisfied, has no influence on the
fidelity of our scheme. Thus the fidelity of our scheme is not
random, and is determined with the driving pulses chosen.
Furthermore, our scheme wins an advantage over that of
Ref.\cite{14} on the high fidelity. Affected by the thermal
oscillation of the atom and the randomness on which well the atom
is trapped in, the average fidelity of the scheme in Ref.\cite{14}
is not as high as that of our scheme.

\section{conclusion}

In summary, we have presented a scheme to teleport the internal state of an
atom trapped in a cavity to another atom trapped in a distant cavity by
measuring interference of polarized photons through single-photon detectors.
Our scheme has a high fidelity of almost unity and a large success
probability. Compared with the original scheme, it has several advantages
including intrinsic robustness to detection inefficiency and randomness in
the atom's position, and thus well fit the status of the current experiment
technology.

\begin{acknowledgments}
We thank L.-M. Duan, Chao Han and Wei Jiang for valuable discussions. This
work was funded by National Fundamental Research Program (2001CB309300),
National Natural Science Foundation of China under Grants No.10204020, the
Innovation funds from Chinese Academy of Sciences.
\end{acknowledgments}

\end{document}